\documentclass[iop]{emulateapj} 
\usepackage{natbib}
\usepackage[normalem]{ulem}
\usepackage{color}

\shorttitle{On the Excitation and Formation of Fullerenes}

\begin{document}

\title{On the Excitation and Formation of Circumstellar Fullerenes}

\author{J.~Bernard-Salas\altaffilmark{1},
  J. Cami\altaffilmark{2,3}, E.~Peeters\altaffilmark{2,3},
  A.P. Jones\altaffilmark{1}, E.R. Micelotta\altaffilmark{2},
  M.A.T. Groenewegen\altaffilmark{4}}

\altaffiltext{1}{Institut d'Astrophysique Spatiale,
  CNRS/Universit\'{e} Paris-Sud 11, 91405 Orsay, France}
\altaffiltext{2}{Department of Physics and Astronomy, The University
  of Western Ontario, London, ON N6A 3K7, Canada}
\altaffiltext{3}{SETI Institute, 189
    Bernardo Avenue, Suite 100,  Mountain View, CA 94043, USA}
\altaffiltext{4}{Royal Observatory of Belgium, Ringlaan 3, 1180
  Brussels, Belgium}

\email{email: jbernard@ias.u-psud.fr}

\begin{abstract}

  We compare and analyze the Spitzer mid-infrared spectrum of
  three fullerene-rich planetary nebulae in the Milky Way and the
  Magellanic Clouds; Tc1, SMP~SMC~16, and SMP~LMC~56. The three
  planetary nebulae share many spectroscopic
  similarities. The strongest circumstellar emission bands
  correspond to the infrared active vibrational modes of the
  fullerene species C$_{60}$ and little or no emission is present from
  Polycyclic Aromatic Hydrocarbons (PAHs). The strength of the
  fullerene bands in the three planetary nebulae is very similar,
  while the ratio of the \ion{[Ne}{3]}15.5\,$\mu$m/
  \ion{[Ne}{2]}12.8\,$\mu$m fine structure lines, an indicator of the
  strength of the radiation field, is markedly different. This raises
  questions about their excitation mechanism and we compare the
    fullerene emission to fluorescent and thermal models. In
  addition, the spectra show other interesting and common features,
  most notably in the 6--9\,$\mu$m region, where a broad plateau with
  substructure dominates the emission.  These features have previously been associated with mixtures of aromatic/aliphatic hydrocarbon solids. We hypothesize on the origin
  of this band, which is likely related to the fullerene formation
  mechanism, and compare it with modeled Hydrogenated Amorphous
  Carbon that present emission in this region.
\end{abstract}


\section{Introduction}
The C$_{60}$ molecule, buckminsterfullerene, was discovered in
laboratory experiments aimed at understanding the formation of long
carbon chains in the circumstellar environment of carbon stars and
their survival in the interstellar medium \citep[ISM,][]{kroto:C60discovery}. In
these experiments, graphite was vaporized in a hydrogen-poor
atmosphere using helium as a buffer gas, resulting in clusters of
carbonaceous particles of different sizes. Amongst the cluster
population, the particles with 60 carbon atoms were the most stable
species and the researchers concluded that these species are
structured like a truncated icosahedron -- often compared to
  the geometry of a black and white soccer ball. Fullerenes are now
known as a class of carbon based molecules in the shape of a
hollow sphere or ellipsoid. Fullerenes can be formed very efficiently
in laboratory experiments, converting a few percent of graphite into
C$_{60}$ \citep{Kraetschmer:UV_IR,Kraetschmer:solidC60}.

As soon as they were discovered, it was suggested that their extreme
stability, in particular against photodissocation, makes fullerenes
ideally suited to survive the harsh radiation field in the ISM, and
thus could be widespread in the galaxy \citep[e.g.][]{kroto:C60discovery}. Once injected into the ISM, 
they could contribute to interstellar extinction, heating, charge
exchange with ions, and provide active surfaces for complex chemical
reactions \citep{KrotoJura,Foing:C60+DIBs}. They have also been
suggested to be responsible for the dust-correlated excess in
microwave background radiation observed in some molecular clouds
\citep{Watson2005,igl04, igl06}.  

Recently, we have detected and identified the vibrational
modes of the fullerene species C$_{60}$ and C$_{70}$ in the
Spitzer-IRS spectrum of the young planetary nebula Tc~1
\citep[][Paper\,I hereafter]{Cami:C60Science}. C$_{60}$ has now been
detected in many more evolved stars: a handful of PNe in the
  Milky Way \citep{Garcia2010:C60_PN_MW} and the Magellanic Clouds
  \citep{Garcia2011:C60_PN_MC}; in the protoplanetary nebula
  \object{IRAS 01005+7910} \citep{Zhang2011}; in the post-AGB stars
  \object{HD~52961}, \object{IRAS~06338+5333} \citep{Gielen:C60-pAGB}
  and \object{HR 4049} \citep{Roberts2011:C60}; in the surroundings of
  a few R~Cor~Bor stars \citep{Clayton2011:RCB,Garcia2011:RCB} and in
  the peculiar binary object \object{XX~Oph} \citep{Evans:C60_XX_Oph}.
  These detections suggest that fullerenes are formed by the complex,
  rich circumstellar chemistry that occurs in the short transition
  phase from AGB star towards PN \citep[Paper I,][]{Zhang2011}, or
  more generally in the circumstellar environments of carbon-rich
  evolved stars. However, it is not clear how the fullerenes form.
Proposed mechanisms include the formation in hydrogen-poor
  environments; photo-chemical processing of Hydrogenated Amorphous
Carbon (a-C:H or HAC) grains; high temperature formation in carbon-rich
environments or the formation of fullerenes from the
destruction of PAHs.

The IR C$_{60}$ bands have also been seen in the interstellar
  medium and in young stellar objects. Following their earlier
  suggestions \citep{Sellgren2007}, \citet{Sellgren:C60_RNe}
confirmed the presence of C$_{60}$ in the reflection nebulae
\object{NGC~7023} and \object{NGC~2023}, and showed that in
NGC7023, the fullerenes emision comes from a different location than
that of the
Polycyclic Aromatic Hydrocarbon (PAH) bands. \citet{Rubin2011}
  furthermore report the detection of C$_{60}$ in the Orion nebula,
  and \citet{Roberts2011:C60} found the C$_{60}$ bands in a few young
  stellar objects and a Herbig Ae/Be star.  These
detections show that fullerenes can survive the
conditions in the ISM and become incorporated into the regions around  young
  stars and, possibly, planetary systems. 

A key question in the studies of fullerenes in astrophysical
  environments is what drives the excitation of these species. This is
  fundamentally important, since it determines how the fullerene bands
  can be used to probe the physical conditions of the environment in
  which they reside. While the presence of circumstellar and
  interstellar fullerenes is now firmly established, many questions
  about their excitation mechanism remain \citep[see][for a
  review]{Cami2011:IAU}. 
Three different mechanisms have been considered. In Paper I, we
  showed that the relative strength of the C$_{60}$ and C$_{70}$ bands
  in Tc~1 are consistent with a thermal distribution over the excited
  states. Such an excitation mechanism could be understood if the
  fullerenes are not free gas-phase species, but are instead in the
  solid state or attached to dust grains. Such a solid state origin
  could also explain the lack of anharmonicities in the band profiles
  and the apparent lack of ionized fullerenes.
  \citet{Sellgren:C60_RNe} and \cite{ber12} on the other hand assume
  that the fullerene emission originates from IR fluorescence of
  isolated free molecules in the gas phase. They compare the band
  ratios of the 7.0\,$\mu$m/18.9\,$\mu$m fullerene bands in their
  observations to Monte Carlo simulations for stochastic heating and
  fluorescent cooling, and find agreement for one object but not the
  other \citep{Sellgren:C60_RNe}. The third mechanism is based on chemical excitation, and involves H atom recombination in HAC materials \citep{dul11}.

In spite of clear spectral differences resulting from both
  mechanisms (see \S~\ref{Sect:Excitation}), there is no clear
  consensus yet on the precise mechanism that operates in the
  astrophysical environments where fullerenes reside, and
  observational support for either mechanism can be found. In several
  cases where the 17.4 and 18.9\,$\mu$m bands are detected, the 7.0 and
  8.5\,$\mu$m bands are very weak or absent, and this is hard to
  understand when considering fluorescence. On the other hand, little
  variation has been reported in the relative band strengths of the
  17.4 and 18.9\,$\mu$m bands, which is hard to understand in the
  framework of thermal models. 

Two contributing factors have hampered progress in
  determining the excitation mechanism. First, most observations that
  exhibit the fullerene bands are strongly affected by PAH
  emission, which makes a good determination of the fullerene band
  strengths very difficult at best. Second, there is a large scatter
  in the existing literature about what are the intrinsic band strengths
  of the fullerene bands. Consequently, the measured
  observational values could lead to quite different conclusions
  depending on what intrinsic values are used.  \\

In this paper we analyze and compare the Spitzer-IRS spectra of
  three PNe exhibiting clear and strong emission of circumstellar
  fullerenes. Two of these PNe, Tc1 and \object{SMP~SMC~16,} have been
  published in the literature, while the mid-IR spectrum of the third
  object, \object{SMP~LMC~56}, is presented here for the first
  time. The three PNe are ideally suited to study the excitation
  mechanism of circumstellar fullerenes: there is no discernible PAH
  emission present that could have a big contaminating influence on the
  fullerene band strengths; the relative strength of the
  fine-structure lines furthermore indicates that the overall
  excitation conditions in the three nebulae are different; and we
  have some additional spatial information for one object (Tc~1). At
  the same time, the three spectra offer intriguing clues about the
  formation of circumstellar fullerenes. 

This paper is organized as follows. In Sect.~\ref{Sect:Data}, we
describe the observations and data reduction steps. Sect.~\ref{Sect:nebulae} summarizes what we know about
the conditions in the three objects. In Sect.~\ref{Sect:Excitation},
we investigate the excitation process by comparing the observed fullerene band strengths to thermal and
fluorescence models using different literature sources for the
intrinsic band strengths. We discuss the formation of fullerenes in
Sect.~\ref{Sect:Formation}.

\section{Data \& Measurements}
\label{Sect:Data}

\subsection{Observations}

\begin{figure}
\includegraphics[angle=90,width=9cm]{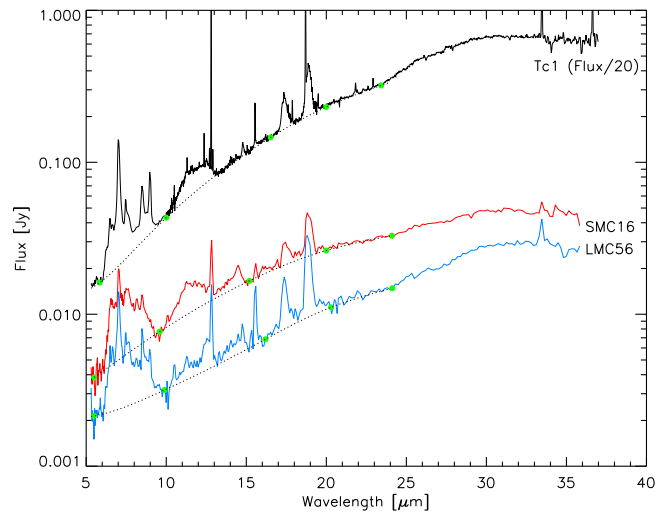}
\caption{\label{Fig:RawSpectra}Observed Spitzer-IRS spectrum of Tc1
  (divided by 20), SMP~LMC~56, and SMP~SMC~16. The dashed line
  indicates the adopted dust continuum, and the green
  dots the anchor points used to define it.}
\end{figure}

The observations presented here were carried out by the
Infrared Spectrograph \citep[IRS,][]{Houck:IRS} on board the
Spitzer Space Telescope
\citep[SST,][]{Werner:Spitzer}. The mid-IR spectrum of Tc1
was published by \citet{per09} and our group (Paper\,I) and consists of
observations at high resolution covering the 10--36\,$\mu$m
  range (using the short-high (SH) and long-high (LH) modules) as well
  as observations using the short-low (SL) and long-low (LL) modules
  covering the wavelength range between 5.4--36\,$\mu$m at a variable
  resolution of 60--120.  SMP~SMC~16 was part of the sample presented
by \citet{Garcia2011:C60_PN_MC} but we have re-extracted its spectrum
with our method for consistency. The spectrum of SMP~LMC~56 is
  obtained from the Spitzer Archive (AOR key $=$ 22423808,
PID$=$40159, PI: A.G.G.M. Tielens). Both SMP~LMC~56 and
SMC~SMC~16 were observed at low resolution only using the SL and
  LL modules.

The data were processed using version S18.7 of the {\em Spitzer}
Science Center's pipeline and, for SMP~LMC~56 and SMP~SMC16, using the new optimal extraction
algorithm in {\em Smart}\footnote{Smart can be downloaded from
  this web site
  http://ssc.spitzer.caltech.edu/archanaly/contributed/smart/index.html.}
\citep{Higdon:Smart,Lebouteiller:OptExt}. The extraction procedures
for these two objects follow those of our earlier papers \citep[Paper
    I;][]{Jero:PN-MC}; we summarize the main steps below.

The data reduction started from the basic calibrated data products
(bcd). First, rogue or unstable pixels were removed and replaced using
the irsclean tool\footnote{This tool is available from the SSC web
  site: http://ssc.spitzer.caltech.edu}.  Then the different cycles
were combined for a given module and order. To remove the background,
the nod positions for a given module and order combination where
subtracted from each other.  There seems to be some extended emission
in SMP~LMC~56 around 33\,$\mu$m in the LL1 module which could affect
the \ion{[S}{3]} 33.4\,$\mu$m line.  Similarly, this spectrum also shows
a slight difference in the absolute flux between nod1 and nod2 in the
LL modules ($\sim$12\%). None of these affect in any way the
conclusions of the paper. The final step is to extract the differenced
images.  Both objects are a point source for the Spitzer-IRS beam and
thus we used the optimal extraction algorithm implemented in {\em
  Smart}. After optimal extraction, any remaining glitches that may
have prevailed from the previous steps were removed manually. Finally,
the two nods were combined to increase the signal-to-noise. No scaling
was needed between the modules. The resulting spectra are shown in
Figure~\ref{Fig:RawSpectra}.

\subsection{The spectra}

\begin{figure*}
\resizebox{\hsize}{!}{\includegraphics[]{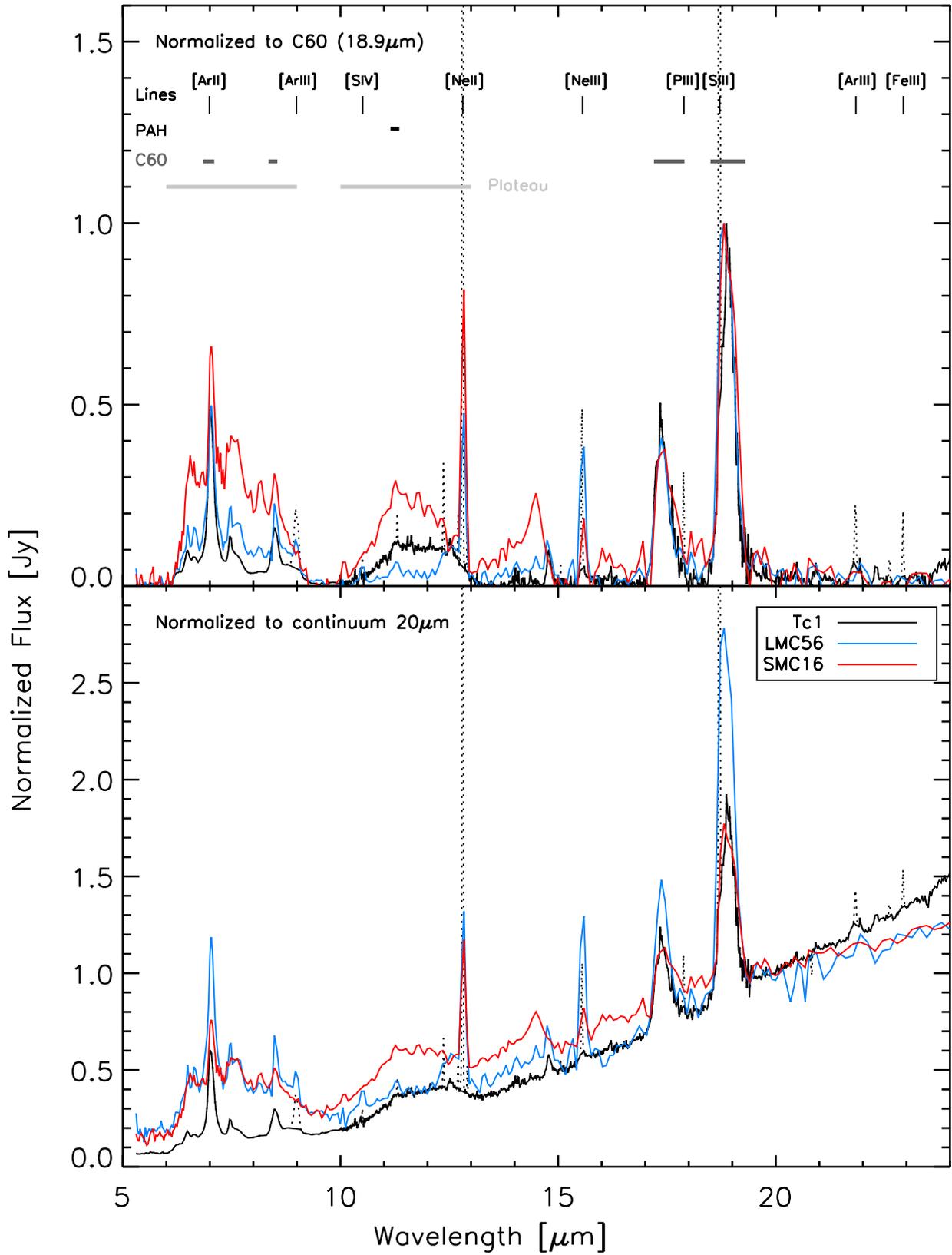}}
\caption{\label{Fig:normspec}Top panel: Continuum subtracted
  spectra of the 5-24\,$\mu$m region normalized to the peak of the
    18.9\,$\mu$m feature. Atomic fine-structure lines in Tc~1 are
  shown as dotted lines. Bottom panel: Spectra normalized to
  the 20\,$\mu$m continuum flux. As can be seen the three PNe
  have very similar continuum emission.  \label{fig2}}
\end{figure*}

To aid further discussions, we provide a brief description here
  of the different features in the spectra.
As can be seen from Fig.~\ref{Fig:RawSpectra}, the spectra of
  Tc1, SMP~LMC~56, and SMP~SMC~16 reveal the same overall shape: a
  strong rising dust continuum on which many emission features are
  superposed. Some of these are better seen in
  Fig.~\ref{Fig:normspec} where we show continuum subtracted, 
  as well as normalized, spectra.

The spectra exhibit many low excitation fine-structure lines.
These show up clearly as narrow lines in the high resolution part
  of the spectrum of Tc 1 ($\lambda \ga 10\, \mu$m) which makes it
  straightforward to distinguish them from other spectral features. In
  the low-resolution observations, the much larger line width makes it
  harder to separate the fine-structure lines from other
  bands. The most prominent lines are [\ion{Ar}{2}]
  (6.99\,$\mu$m), [\ion{Ar}{3}] (8.99\,$\mu$m), [\ion{Ne}{2}]
  (12.81\,$\mu$m), [\ion{Ne}{3}] (15.55\,$\mu$m) and [\ion{S}{3}] (18.7
 \,$\mu$m). A weak \ion{[S}{4]} line at 10.51\,$\mu$m is present as
  well. In Tc~1, the \ion{[S}{3]} line at 18.7\,$\mu$m is perched on
  top of the much broader C$_{60}$\ band. In the low-resolution spectra of SMP~LMC~56 and SMP~SMC~16, the
    presence of the [\ion{S}{3}] can be inferred from band asymmetries
    (see Appendix). We note that the
    34.8\,$\mu$m line of \ion{[S}{3]} in SMP~LMC~56 is contaminated by
    the extended emission mentioned in \S2.1, and we cannot determine
    how much of it originates from the actual source.

These spectra furthermore clearly show fullerene emission
  bands, and it is remarkable how similar the relative strength of the
  fullerene bands are in these three PNe (Fig. 2 top). There is little to no
  contamination by PAH features, and thus these three objects
  represent some of the clearest detections of fullerenes. Most
  notable are the strong and broad C$_{60}$ bands at 7.0, 8.5, 17.4
  and 18.9\,$\mu$m. Tc1 furthermore also exhibits weaker
  features at 12.7, 14.8, 15.6 and 21.8\,$\mu$m that are attributed to
  $C_{70}$ (Paper\,I); C$_{70}$ also contributes ($\sim$ 10\% of the
  total power) to the 17.4 and 18.9\,$\mu$m bands. These features could
  be present in the spectrum of SMP~LMC~56, but seem absent in
  SMP~SMC~16.

Individually, the spectra show little evidence for the presence of
PAHs, which are typically seen in carbon-rich PNe. However, the
three spectra do show a weak bump at 11.3\,$\mu$m; emission
  at these wavelengths is generally attributed to PAHs. SMP~LMC~56 also shows a feature  near
  12.7\,$\mu$m where another PAH band is commonly seen; however,
  in our case, this feature may be due to C$_{70}$. Other PAH bands
are not easily identified. Notably absent is the usually
  strong 7.7\,$\mu$m PAH band. There is a very weak bump near 6.2\,$\mu$m
in Tc1 and SMP~LMC~56, which is not clear or present in
SMP~SMC~16 while any possible 8.6\,$\mu$m PAH emission will be
blended with the C$_{60}$ feature. 

Very striking in the spectra of all three PNe are the broad
  emission plateaus between 6--9\,$\mu$m and between 10--13\,$\mu$m. 
 Similar plateaus in these spectral regions have been attributed to modes of alkane and alkene groups on the periphery of polycyclic aromatic systems in proto-PNe  by \citet{kwo01}. 
The 6--9\,$\mu$m plateau is  seen in very few other PNe and could thus
  well be related to the fullerene formation process. In fact,
\citet{Garcia2010:C60_PN_MW} hypothesize that these features are
due to HACs, and that fullerenes are formed from the
  decomposition of these HAC grains.  We discuss this further
in \S5. In this paper we use the term HAC in its broader sense to include the whole family of these materials (e.g. petroleum, coals, quenched carbonaceous composite a-C:H, a-C, etcetera).

The 6-9\,$\mu$m plateau reveals some substructure as well. This substructure is present in both nod positions, and 
  in each of the three objects confirming that these are  real spectral
  features. Tc 1 and SMP~LMC~56 have features at 6.49 and 6.65
$\mu$m. These are also observed in a few post-AGB stars
\citep{Gielen:C60-pAGB}. In addition, there is a fairly broad
feature between roughly 7.35 and 7.85\,$\mu$m best seen in
  SMP SMC 16 and SMP LMC 56. The width of the feature is about the
same as that of the C$_{60}$ bands. It is interesting to
  note that 7.5\,$\mu$m is about the wavelength where a C$_{60}^{+}$
  band is expected \citep{ful93}. Tc1 and SMP~LMC~56 also show
a small peak at about 7.48\,$\mu$m, most likely due to the \ion{H}{1}
recombination line. Finally, SMP~LMC~56 and SMP~SMC~16 show a
feature at 8.15\,$\mu$m which is not present in Tc1. 
  
The plateau emission in the 10-13\,$\mu$m region is quite strong
in Tc~1 and SMP~SMC~16, and weaker in SMP~LMC~56. A 10-13\,$\mu$m
  emission plateau is also observed towards other PNe
  \citep{Jero:PN-MC} and is generally attributed to SiC
  \citep{Speck2005:SiC,Speck2009:SiC}. However, the profile of the
  plateau in our three PNe is quite different from those PNe, and
  suggests that here, a different carrier might be responsible.  

In all three sources, we also detect the so-called 30\,$\mu$m
  feature that is commonly seen in carbon-rich PNe. This feature is
  often attributed to MgS \citep{Hony:MgS,Jero:PN-MC}, but could also
  have a carbonaceous origin \citep[e.g.,][]{vol11}. Finally, SMC SMC 16 shows an asymmetric feature that peaks at
14.5\,$\mu$m; it is not clear what the origin is of this feature.

\subsection{Flux Measurements}

Detail information on how the different features (continuum, atomic lines, fullerene bands) were measured is given in the Appendix, and these values are listed in Table 1.

\begin{deluxetable*}{llrrr}
  \tablecaption{\label{Table:Fluxes}Flux Measurements -- all numbers in W/m$^2$}
  \tablehead{  & & \multicolumn{1}{c}{Tc 1} & \multicolumn{1}{c}{SMC 16} & \multicolumn{1}{c}{LMC 56}}\\
  \startdata 

  \multicolumn{2}{l}{$F_{\rm dust}$} & 6.0 $\times$ 10$^{-13}$ 
                                  & 4.5 $\times 10^{-15}$ 
                                  & 1.9 $\times 10^{-15}$\\

  \multicolumn{2}{l}{[\ion{Ne}{2}] (12.8\,$\mu$m)} & 2.3 $\times 10^{-14}$
                                    & 3.5 $\times 10^{-17}$
                                    & 2.1 $\times 10^{-17}$\\
  \multicolumn{2}{l}{[\ion{Ne}{3}] (15.5\,$\mu$m)} & 8.9 $\times 10^{-16}$
                                    & 7.7 $\times 10^{-18}$
                                    & 1.8 $\times 10^{-17}$\\

  \multicolumn{2}{l}{18.9\,$\mu$m band}\\

  & Total\tablenotemark{a}  & 2.4 $\times 10^{-14}$ 
                            & 9.5 $\times 10^{-17}$
                            & 9.3 $\times 10^{-17}$ \\

  & [\ion{S}{3}]\tablenotemark{b} & 8.7 $\times 10^{-15}$ 
                                  & 3.4 $\times 10^{-17}$
                                  & 3.4 $\times 10^{-17}$ \\

  & Fullerenes\tablenotemark{b} & 1.4 $\times 10^{-14}$ 
                                & 4.1 $\times 10^{-17}$
                                & 4.4 $\times 10^{-17}$ \\

  & Baseline\tablenotemark{b}   & 2.0 $\times 10^{-15}$ 
                                & 2.0 $\times 10^{-17}$
                                & 1.5 $\times 10^{-17}$\\

  & C$_{70}$\tablenotemark{c}   & $\sim$ 1.4 $\times 10^{-15}$ 
                               & 
                               &  \\

  & C$_{60}$                    & $1.3 \times 10^{-14}$ 
                               & 5.1 $\times 10^{-17}$
                               & 5.2 $\times 10^{-17}$ \\

  & Uncertainty                & 1 $\times 10^{-15}$ 
                               & 1 $\times 10^{-17}$
                               & 8 $\times 10^{-18}$\\

  \multicolumn{2}{l}{17.4\,$\mu$m band} \\
  & Total\tablenotemark{a} & 9.2 $\times 10^{-15}$ 
                           & 4.3 $\times 10^{-17}$
                           & 4.3 $\times 10^{-17}$\\  

  & Contaminants\tablenotemark{e} & 6.1 $\times 10^{-16}$ \\

  & Baseline                      & 1.2 $\times 10^{-15}$ 
                                  & 1.3 $\times 10^{-17}$
                                  & 1.6 $\times 10^{-17}$\\

  & C$_{70}$\tablenotemark{c}   & $\sim 1.8 \times 10^{-15}$ \\

  & C$_{60}$                    & 6.2 $\times 10^{-15}$ 
                               & 3.7 $\times 10^{-17}$
                               & 3.5 $\times 10^{-17}$\\

  & Uncertainty                & 6 $\times 10^{-16}$ 
                               & 7 $\times 10^{-18}$
                               & 8 $\times 10^{-18}$\\

  \multicolumn{2}{l}{8.5\,$\mu$m band} \\
  & C$_{60}$                    & 3.9 $\times 10^{-15}$ 
                               & 1.8 $\times 10^{-17}$
                               & 1.3 $\times 10^{-17}$ \\

  & Uncertainty                & 2 $\times 10^{-16}$
                               & 3 $\times 10^{-18}$
                               & 3 $\times 10^{-18}$\\

  \multicolumn{2}{l}{7.0\,$\mu$m band} \\
  & Total\tablenotemark{a} & 2.1 $\times 10^{-14}$ 
                           & 6.8 $\times 10^{-17}$ 
                           & 9.0 $\times 10^{-17}$ \\

  & [\ion{Ar}{2}]\tablenotemark{b} & 3.4 $\times 10^{-15}$ 
                                   & 2.5 $\times 10^{-17}$ 
                                   & \multicolumn{1}{c}{--}\\

  & C$_{60}$                        & 1.7 $\times 10^{-14}$ 
                                   & 4.1 $\times 10^{-17}$ 
                                   & 9.0 $\times 10^{-17}$ \\

  & Uncertainty                    & 3 $\times 10^{-15}$ 
                                   & 1 $\times 10^{-17}$
                                   & 1 $\times 10^{-17}$\\
  \multicolumn{2}{l}{F(17.4$\mu$m)/F(18.9$\mu$m)} & 0.47 $\pm$ 0.06
                                   & 0.72 $\pm$ 0.29
                                   & 0.68 $\pm$ 0.18\\
  \multicolumn{2}{l}{F(8.5$\mu$m)/F(18.9$\mu$m)} & 0.29 $\pm$ 0.02 
                                   & 0.36 $\pm$ 0.09 
                                   & 0.26 $\pm$ 0.05 \\
  \multicolumn{2}{l}{F(7.0$\mu$m)/F(18.9$\mu$m)} & 1.31 $\pm$ 0.28 
                                   & 0.80 $\pm$ 0.42 
                                   & 1.74 $\pm$ 0.63 \\

\enddata
\tablenotetext{a}{From integrating over the entire band. }
\tablenotetext{b}{From fitting the observations with two Gaussian
  profiles and a linear baseline. }
\tablenotetext{c}{Estimate from Paper I.} 
\tablenotetext{d}{Mainly due to misspossitioning continuum. } 
\tablenotetext{e}{For Tc~1: the [\ion{P}{3}] line and a weaker line at
  17.6\,$\mu$m.}

\end{deluxetable*}

For what follows, the most important quantities are the flux ratios of
the different C$_{60}$ bands. We have included in
Table~\ref{Table:Fluxes} the band ratios normalized to the 18.9\,$\mu$m
band. It is immediately clear that the derived $F$(8.5\,$\mu$m) /
$F$(18.9\,$\mu$m) ratios are very similar, and are in fact compatible
with a constant ratio equal to the weighted mean value of
0.29~$\pm$~0.02. The $F$(17.4\,$\mu$m) / $F$(18.9\,$\mu$m) ratio shows a
somewhat larger spread but also larger uncertainties. The weighted
mean value is 0.50~$\pm$~0.06, and all measurements are thus
consistent with a constant value. For the ratios involving the 7
$\mu$m band, there is a much larger spread in the resulting band
ratios which stems to a large degree from the difficulties in
determining the contribution of the [\ion{Ar}{2}] line to the 7\,$\mu$m
emission. The weighted mean ratio is 1.22~$\pm$0.22, and given the
uncertainties, it is not clear whether any real variations are
present.

In Table~1 we furthermore list the fluxes of the
\ion{[Ne}{2}] (12.8\,$\mu$m) and \ion{[Ne}{3]} (15.5\,$\mu$m),
  fine-structure lines that will be of importance later on.

\subsection{Spatial distribution}
\label{Sect:spatial}

 \begin{figure}
\includegraphics[width=9.cm]{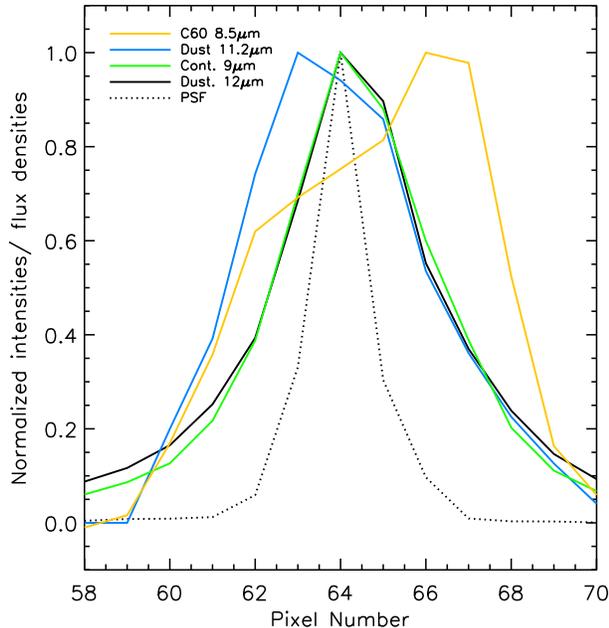}
\caption{ The spatial distribution of different dust
  components in Tc1 along the SL slit. The point spread function
    (PSF) is shown for reference.  The central star is located at
    pixel 64. All components are extended; but whereas the dust
    continuum emission is roughly centered on the star, the fullerene
    emission peaks 2-3 pixels away (to the right in this figure) from
    the central star. The 11.2\,$\mu$m feature peaks about a pixels away
    from the central star on the other side of the fullerene
    emission.\label{Fig:Spatial}}
\end{figure}

Because of its proximity and size ($\sim$9$\arcsec$), we can study the
spatial distribution of several emission components in Tc~1
relative to the position of the central star in the SL slit
(1.8$\arcsec$ per pixel in the spatial direction, Figure ~3);
unfortunately, this is not possible for the high-resolution
  observations nor for the much more distant SMP SMC 16 and SMP LMC
  56. We mapped the distribution of the thermally emitting dust
  by using the flux at 9.46\,$\mu$m. From Fig.~3, it is clear
  that the dust emission is extended and centered on the central
  star. The same holds true for the flux at 12\,$\mu$m, which traces
  the dust continuum in addition to the broad 10--13\,$\mu$m plateau
  (see Fig.~\ref{Fig:normspec}). The emission in the 8.5\,$\mu$m
  fullerene band and in the weak 11.2\,$\mu$m feature is even more
  extended. However, in both cases, the emission is displaced
  from the central star and the two different components peak at
opposite direction from the central star. This is reminiscent of
  the reflection nebulae NGC\,7023 and NGC\,2023 where \citet{Sellgren:C60_RNe} and \citet{pee12}
  reported a similar spatial separation in the distribution of
  fullerene and PAH emission.The 2-3 pixel displacement of the
  fullerene emission corresponds to ~6400-9700 AU at the distance of
  Tc~1 \citep[1.8 kpc, ][]{Pottasch2011}.

\section{Nebular Conditions}
\label{Sect:nebulae}

As hinted at by their spectroscopic resemblance, Tc1,
SMP~LMC~56, and SMP~SMC~16 share some physical properties. The observed line emission is dominated by low excitation lines, typical a of low excitation PNe. The dust emissions starts to rise $<$5\,$\mu$m indicating the presence of a hot dust component. This continuum emission is typical and similar to other young PNe like NGC7027 and BD+30\,3639 \citep{ber03,ber05}.
Further evidence for a young age is the fact that Tc1 and
SMP~SMC~16 are relatively compact objects. Furthermore, the effective temperatures (T$_{eff}$) of the central stars are low. Tc1 has a T$_{eff}$ of 34\,700\,K and SMP~SMC~16 of 37\,000\,K \citep{dop91,Pottasch2011}. \citet{Villaver2003} note that the Zanstra temperature in SMP~LMC~56
is low (45 900 and 29 000\,K for He\,II and H\,I respectively), and
places the object in the early stages of the Helium burning tracks of
\citet{VassiliadisWood1994}, again indicating that it is
young. The electron temperatures are:  9\,000\,K for Tc1 \citep{Pottasch2011},
$\sim$13\,100\,K and 11\,800 for SMP~LMC~56 
  SMP~SMC~16 respectively \citep{Leisy2006}. These are  are not very high when compared to other PNe in their host galaxies \citep[see e.g. samples
by][]{Stanghellini2007,Jero:PN-MC}.

Tc1 and SMP~SMC~16 are carbon-rich, with high C/O ratios of 1.4 and 2.1
respectively
\citep[from abundances by][]{Pottasch2011,Stanghellini2009,Leisy2006}. There are no carbon
abundances available in the literature for SMP~LMC~56 but
its mid-IR spectral characteristics are clearly that of a
carbon-rich environment (e.g. presence of fullerenes, 30\,$\mu$m
feature). As it has been suggested before, these conditions (young, carbon-rich, low excitation) appears to favor the formation of fullerenes.

Although the conditions in the three PNe are quite similar, they
  are not the same. In particular, the observed line emission suggests
  that the overall excitation conditions are somewhat different. This
  follows from the [\ion{Ne}{3}]/[\ion{Ne}{2}] ratio which is an
  indicator for the strength of the radiation field and which changes
  significantly -- from about 0.04 in Tc~1 to 0.22 in SMP~SMC~16 and
  even 0.86 in SMP~LMC~56.

\section{The excitation of circumstellar fullerenes}
\label{Sect:Excitation}

\subsection{Thermal and fluorescence models}

We start our analysis from the {\em intrinsic} relative strengths
  for the C$_{60}$ bands. The literature offers quite a few sources
  where band strengths have been obtained theoretically or
  experimentally, and there is a fair amount of disagreement in the
  relative strengths of the C$_{60}$ bands. Here, we have chosen four
  different literature sources 
  \citep{Chase1992,Fabian1996,Choi_etal_2000,Iglesias:C60strengths}
  from both theoretical and experimental results which together are
  fairly representative of the band strength differences found in the
  literature. The band strengths (relative to 18.9\,$\mu$m band) are
  listed in Table~\ref{Table:lab_models}. All sources agree that the
  18.9\,$\mu$m band is the strongest, and that the 8.5\,$\mu$m band is
  the weakest, but there is disagreement about the rest, with
  differences in the actual listed values of up to a factor 2.5 (for
  the 17.4\,$\mu$m band). 

\begin{figure*}
\includegraphics[width=12cm]{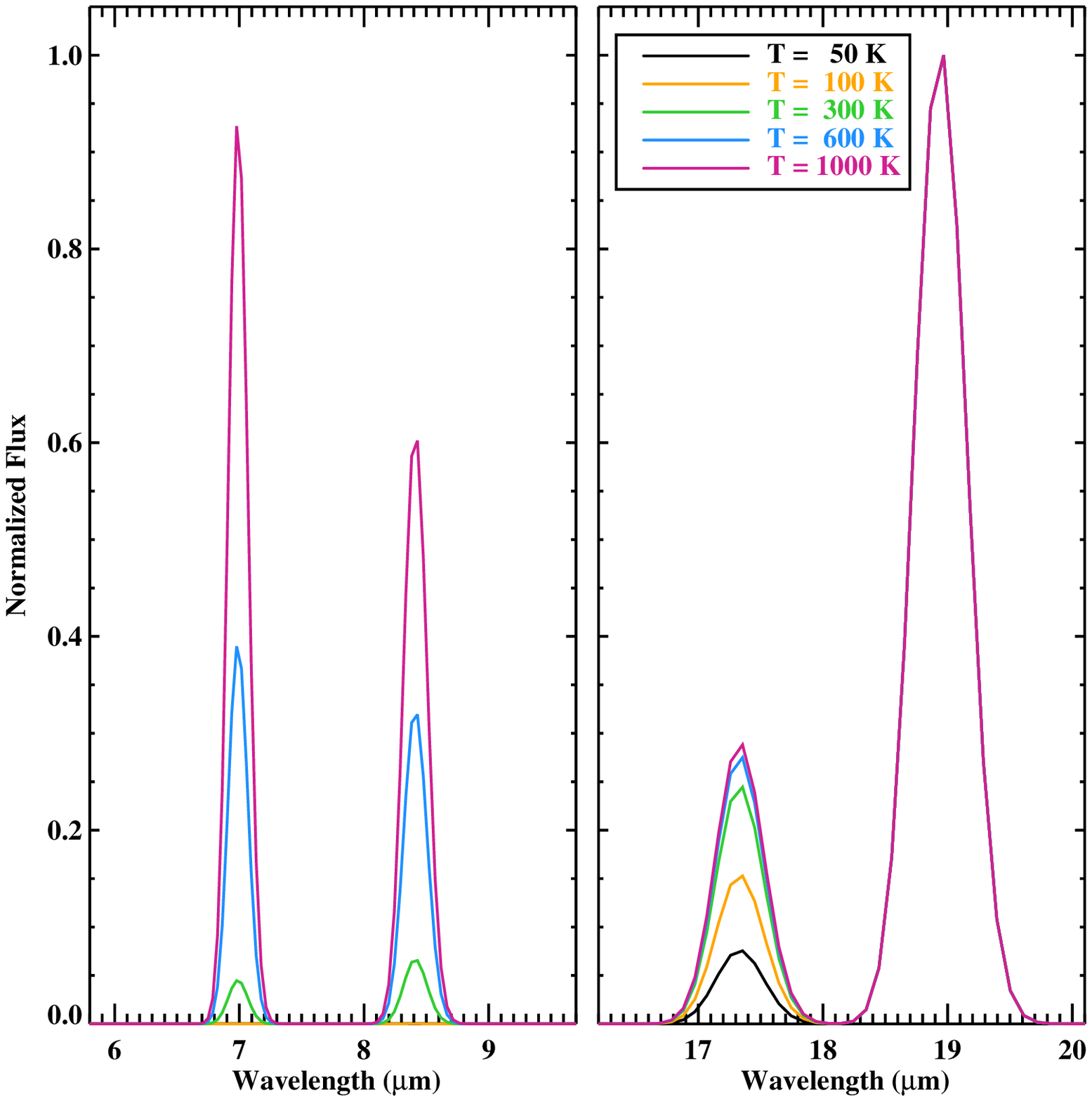}\\
\includegraphics[width=12cm]{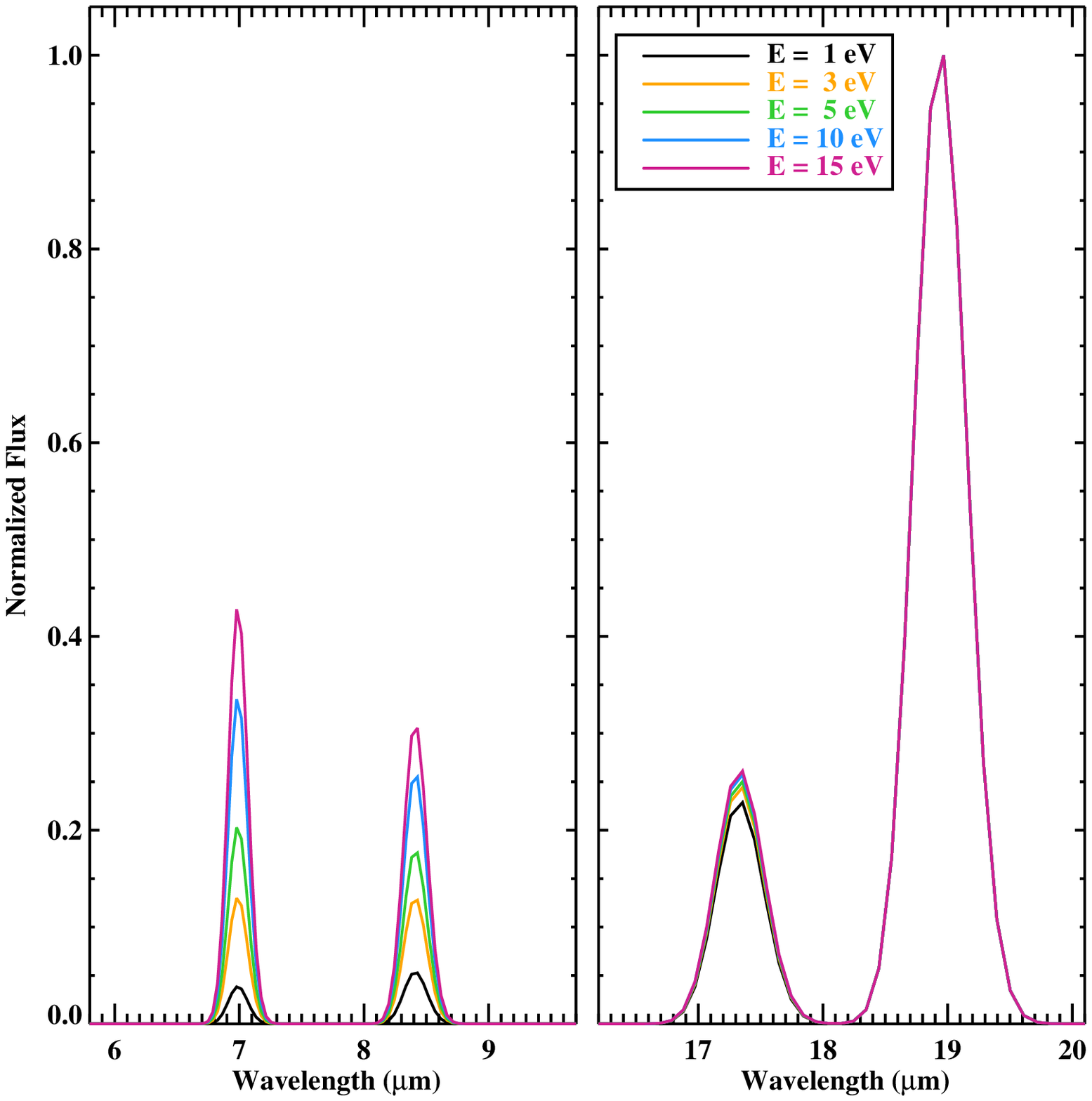}
\caption{\label{Fig:ModelVariations}This figure shows how the relative
  strengths of the C$_{60}$ bands changes as a function of temperature
  for thermal models (top) and as a function of the average absorbed
  photon energy for IR fluorescence models (bottom). In all cases, we
  used the intrinsic band strengths from \citet{Choi_etal_2000}, and
  distributed the total power over Gaussian profiles whose width is
  set by the resolution (taken to be $R=90$). All figures are
  normalized to the peak emission in the 18.9\,$\mu$m band. Note that
  in the observations presented here, the bands at 17.4 and and
  18.9\,$\mu$m are much broader than the instrumental resolution and
  thus these figures cannot be directly compared to the
  observations. We did not include temperature-dependent broadening or
  shifting of the bands. }
\end{figure*}

In the thermal models considered so far, the emitted power in
  each of the C$_{60}$ bands is proportional to the population in the
  corresponding excited vibrational state, and this in turn is set by
  the temperature through the Boltzmann equation. These models thus
  assume that the emission is optically thin, and that emission from
  hot bands can be neglected. Certainly for high temperatures, this
  latter approximation is questionable. Thermal models show that the
  strength of the 17.4\,$\mu$m band (relative to the 18.9\,$\mu$m band)
  changes significantly as a function of temperature for $T\la 500 K$,
  and furthermore that the 7.0 and 8.4\,$\mu$m bands are weak or absent
  for $T\la 300 K$ (see Fig.~\ref{Fig:ModelVariations}). 

The starting point for IR fluorescence models on the other hand is
  the absorption of a single UV photon of energy $\varphi$ that
  results in an electronic transition. On ultra-short timescales, the
  absorbed energy is redistributed over the vibrational modes, which
  causes the molecule to relax back into the electronic ground state,
  but leaves it in highly excited vibrational states. The molecule
  then cools by emitting IR photons as it cascades down the energy
  ladder. The resulting IR spectrum can be calculated in different
  ways; we follow the approach that is used for calculating the IR
  fluorescence of PAHs \citep[see e.g.][]{1989ApJS...71..733A, dHendecourt:thermalapprox,2001ApJ...556..501B,PAHdb}. The
  resulting model spectra are equivalent to those obtained e.g. by a
  Monte Carlo method \citep{Joblin:MC,Sellgren:C60_RNe}. Fluorescence
  models result in a nearly constant F(17.4\,$\mu$m)/F(18.9\,$\mu$m) band
  strength ratio, even for photon energies as low as 1~eV; for
  realistic values of $\varphi$ (5--10 eV), the 7.0 and 8.5\,$\mu$m
  bands are furthermore fairly strong compared to 18.9\,$\mu$m band
  (see Fig.~\ref{Fig:ModelVariations}). 

\citet{dul11} proposed that HAC particles can be heated by the recombination
 of trapped (physi-sorbed) H atoms with dangling bonds in the HAC
 structure. The release of chemical energy upon bond formation heats
 the dust and provides an additional means of exciting the UIR bands.
 Given that the HAC/a-C:H dust in these
 regions is rather warm (~100--150 K), any
 interstitial H atoms may not be retained within the material. In the following section we consider the thermal and fluorescence mechanisms.

\subsection{Models versus observations}

\begin{figure}
\resizebox{\hsize}{!}{\includegraphics{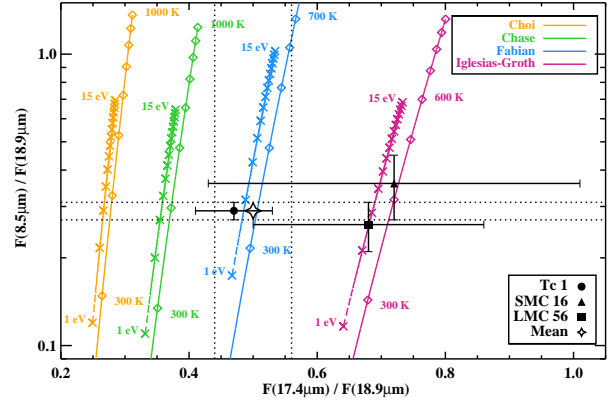}}
\caption{\label{Fig:Diagplot8.5_17.4}A diagnostic plot showing measurements and
  different model values for the total flux in the 8.5\,$\mu$m band and
  the flux in the 17.4\,$\mu$m band (both normalized to the 18.9\,$\mu$m
  band). Each pair of curves corresponds to a different source of
  C$_{60}$ band strengths. The right hand curve for each pair
  corresponds to the band ratios for thermal models. The diamond plot
  symbols correspond to temperature increments of 100 K, and a low and
  a high temperature value is indicated on each curve. The left hand
  curves are model values for IR fluorescence. The crosses correspond
  to increments in the average absorbed photon energy of 1 eV. For
  each curve, the 1 eV and 15 eV models are indicated. We also show
  the measured band ratios for each of our three PNe, and the weighted
  mean value of the band ratios. The dotted lines indicate the
  1$\sigma$ confidence intervals for the mean values. }
\end{figure}

For each set of intrinsic band strengths listed in
  Table~\ref{Table:lab_models}, we have calculated thermal models (for
  temperatures between 50\,K and 1000\,K) and fluorescence models (for
  photon energies between 1 and 15 eV) and calculated the band strength
  ratios for each of those models. Fig.~\ref{Fig:Diagplot8.5_17.4}
  shows the locus of expected strengths of the 8.5\,$\mu$m and 17.4
 \,$\mu$m bands (normalized to the 18.9\,$\mu$m band) for all models,
  and compares these to the observed values for all three PNe (and
  their weighted mean). Since the lowest point on the thermal model
  curves in the diagram corresponds to 300\,K, there is little
  variation left in the 17.4\,$\mu$m band strength for higher
  temperatures, and the thermal and fluorescence models almost
  overlap. Based on the strength of the 17.4\,$\mu$m band, it would
  thus be impossible to differentiate between thermal and fluorescence
  models for this range of parameters. Note also how far apart the
  four different sets of models are, resulting from the very different
  intrinsic strengths of the 17.4\,$\mu$m band in the four sets. 

\begin{deluxetable}{llrrr}
  \tablecaption{\label{Table:lab_models}Intrinsic and observational
    values for the band strengths relative to the $T_{1u}(1)$ band at
    18.9\,$\mu$m; derived temperature ranges and ranges in photon
    energies. } 

  \tablehead{\multicolumn{2}{l}{ } & T$_{1u}(2)$ &
    T$_{1u}(3)$ & T$_{1u}(4)$ \\
    & & (17.4\,$\mu$m) & (8.5\,$\mu$m) & (7.0\,$\mu$m) \\ } \tablewidth{0pt}
  \startdata
  \multicolumn{5}{c}{RELATIVE BAND STRENGTH} \\
  \multicolumn{5}{l}{Intrinsic (literature)} \\
  & Choi           & 0.26 & 0.31 & 0.46 \\
  & Chase          & 0.34 & 0.28 & 0.34 \\
  & Fabian         & 0.48 & 0.45 & 0.38 \\
  & Iglesias-Groth & 0.66 & 0.30 & 0.50 \\
  & \\
  \multicolumn{2}{l}{Observational (mean)} & 0.50 &
  0.29 & 1.22\\
  \multicolumn{2}{l}{ } & $\pm$0.06 &
  $\pm$0.02 & $\pm$0.22\\
  & \\
  \multicolumn{5}{c}{DERIVED PARAMETERS} \\
  \multicolumn{5}{l}{Thermal Models: 1 $\sigma$ temperature ranges (K)} \\
  &Choi           & --         & 370--390 & 585--700\\
& Chase          & --         & 385--405 & 680--840 \\ 
& Fabian         & 200--630 & 320--340  & 645--785 \\ 
& Iglesias-Groth & 105--165 & 375--395 & 565--670\\ 
& \\
\multicolumn{5}{l}{Fluorescence Models: 1 $\sigma$ photon energies (eV)} \\
& Choi           &  --        & 2.7--3.3 & 11.7--22.8 \\
& Chase          &  --        & 3.0--3.7 & 19.0--44.9\\ 
& Fabian         & 0.1--50  & 1.6--2.0  & 15.9--34.6\\ 
& Iglesias-Groth & --         & 2.8--3.4  & 9.9--18.5 \\ 
\enddata
\end{deluxetable}

\begin{figure}
\resizebox{\hsize}{!}{\includegraphics{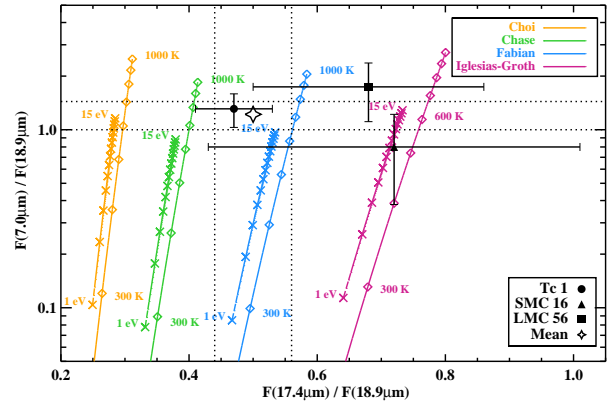}}
\caption{\label{Fig:Diagplot7.0_17.4}Same as
  Fig.~\ref{Fig:Diagplot8.5_17.4}, but for the 7.0
 \,$\mu$m band instead of the 8.5\,$\mu$m band. }
\end{figure}

The 8.5\,$\mu$m band has more diagnostic power. From
  Fig.~\ref{Fig:Diagplot8.5_17.4} it is clear that already at 
  unrealistically low photon energies of 1 eV (the lowest point on the
  curves), the fluorescence models predict a band at 8.5\,$\mu$m that
  is about 10--20\% of the 18.9\,$\mu$m band. For more reasonable
  values of the absorbed energy (5--10 eV), the 8.5\,$\mu$m band should
  be about 40--60\% of the 18.9\,$\mu$m flux. However, the measured
  ratio is $0.29 \pm 0.02$ which then implies a low photon energy
  ($\varphi \approx 3$ eV). In fact, for Tc~1 (and the weighted mean
  value), we can exclude excitation by 5~eV photons at the 5$\sigma$
  (or better) confidence level, independently of the chosen source of
  intrinsic band strengths. Such low photon energies would be very
  hard to understand since it would imply that the fullerenes do not
  get excited through their strong electronic transitions at UV
  wavelengths in spite of a strong radiation field. Since this result
  holds for all sources of intrinsic strength, and since the 8.5
 \,$\mu$m band strength is well determined (i.e. it is not possible
  that we miss a large fraction of the flux in this band), the 8.5
 \,$\mu$m band strength thus excludes quite strongly the possibility
  that the observed fullerene emission is due to fluorescence by
  isolated molecules in the gas phase. Note that for the 7.0\,$\mu$m
  band (Fig.~\ref{Fig:Diagplot7.0_17.4}), the same conclusion holds,
  but then for opposite reasons: explaining the observed 7.0\,$\mu$m
  band strength requires unreasonably high energies for the exciting
  photons. 

Thermal models on the other hand require a temperature in the
  range 300-400\,K to explain the 8.5\,$\mu$m band strength which at
  first sounds reasonable. However, since the relative strength of the
  8.5\,$\mu$m band is nearly the same in all three objects, this
  implies that also the fullerene temperatures are nearly exactly the
  same in all three sources; this would be either very coincidental,
  or an important factor in explaining the presence of
  fullerenes. Furthermore, the observed 7.0\,$\mu$m flux is much higher
  than would be expected for thermal excitation, and this holds for
  all sources of band strength. Thus, the observations are only
  consistent with thermal models if the 7.0\,$\mu$m band would be
  primarily due to [\ion{Ar}{2}] and only a small fraction of the 7.0
 \,$\mu$m flux could be attributed to C$_{60}$. In such a case, the
  intrinsic strengths listed by \citet{Fabian1996} are fully
  consistent with the observations (which were the values we used in
  Paper I). However, we
  do not believe this to be the case (see Appendix) -- most of the 7.0\,$\mu$m seems
  to be originating from C$_{60}$. Thus, we have to conclude that
  neither thermal nor fluorescence models can explain the observations
  completely.

\subsection{Discussion}

In spite of the difficulties in reconciling the observations with
  models, the three spectra presented here offer important clues to
  the excitation mechanism. An important ingredient in understanding
  the excitation of fullerenes in our sources is the spatial
  distribution of some of the spectral components in Tc~1
  (Sect.~\ref{Sect:spatial} and Fig.~\ref{Fig:Spatial}). As we
  discussed, the C$_{60}$ emission peaks at a distance of about
  8000~AU from the central star. Given the central star's effective
  temperature of $T_{\rm eff} \approx 30,000$\,K, classical "graphitic" dust in equilibrium
  with the radiation field would be very cold, with temperatures of
  only 20--25\,K. However, we find that hydrogen-rich HAC nanoparticles can be maintained at temperatures of 100--150 K at distances  of 3000 to 10\,000 AU because of their low emissivities at far-infrared wavelengths. This is still not sufficient to heat the fullerenes to the required thermal temperatures of $\sim$300~K. The spatial information thus rules out thermal emission, and
  favours fluorescence. For fluorescence, the average photon energy
  $\varphi$ absorbed by the molecules is a function of the local
  radiation field and the absorption cross section of the molecule;
  this then quite naturally explains why the fullerene band strengths
  are so similar in the three objects since their central stars have
  comparable effective temperatures and thus the average absorbed
  photon energy is very similar as well; for thermal models such a
  coincidence would required fine-tuned conditions in all three PNe. 

However, as we discussed above, the observations are not
  consistent with models for fluorescent IR emission either: for
  realistic energies, all models for fluorescence predict a much
  stronger 8.5\,$\mu$m band and a much weaker 7.0\,$\mu$m band (all
  relative to the 18.9\,$\mu$m band) than is observed, and this is the
  case for all sources of intrinsic band strength. Note that the
  normalization to the 18.9\,$\mu$m band in itself is not the issue
  either -- if the 18.9\,$\mu$m band would be weaker than we measure,
  this could conceivably bring the 8.5\,$\mu$m band flux in agreement
  with models, but then the problem with the 7.0\,$\mu$m band would
  even be worse. 

Since it is clear that we are not seeing emission from isolated,
  free C$_{60}$ molecules, an obvious first question is whether we're maybe
  seeing emission from fullerene clusters or nano-crystals. Such
  entities that are larger than single molecules could still undergo
  single-photon heating and IR fluorescence through the same
  vibrational modes, but their larger heat capacity would result in
  different band ratios -- with lower fluxes at shorter
  wavelengths. This could possibly explain the discrepancy for the 8.5
 \,$\mu$m band, but not for the 7.0\,$\mu$m band, where the problem
  would be even bigger. 

It is instructive to  look at the issue starting from the 8.5
 \,$\mu$m band, since this band does not seem contaminated by other
  emission components. The problem is then that both the observed 18.9
 \,$\mu$m and the 7.0\,$\mu$m bands are too strong to be explained by
  fluorescence of free isolated C$_{60}$ molecules. This then
  indicates that other materials are contributing to the 7.0 and 18.9
 \,$\mu$m band. For the 18.9\,$\mu$m band we have already identified one
  contributor: C$_{70}$. Given the strength of the other C$_{70}$
  bands in the same wavelength range (for Tc~1), it seems unlikely
  though that we severely underestimate its contribution to the 18.9
 \,$\mu$m band; to explain the discrepancies, we would require roughly
  four times more C$_{70}$ than we have currently estimated. For the
  7.0\,$\mu$m band on the other hand, we have not considered a possible
  contribution of C$_{70}$ yet. The intrinsically strongest C$_{70}$
  band occurs indeed near 7.0\,$\mu$m. When using thermal models at the
  fairly low temperatures we derived in paper I, the transitions at
  the shorter wavelengths are suppressed significantly, to the point
  where this band becomes unmeasurably weak. When using fluorescence
  models however, this is not the case, and based on the intrinsic
  strength of the band, it could well be responsible for a large
  fraction of the observed discrepancy. Note that C$_{70}$ still has
  other features at wavelengths between 5--10\,$\mu$m that should then
  be reconsidered in the framework of fluorescence, and it is not clear
  whether attributing the 7.0\,$\mu$m flux to C$_{70}$ is consistent
  with the expected strengths of these features.

\section{HACs and the formation of fullerenes}
\label{Sect:Formation}

The three spectra we present here also offer intriguing clues to
  the formation of circumstellar fullerenes. Indeed, as we discussed
  in Sect.~\ref{Sect:Data}, the three PNe share other notable spectral
  features in addition to the fullerene bands. Of particular
  importance for the discussion on fullerene formation mechanisms are
  the plateaus.  Several formation routes are possible in the laboratory or have been proposed, including: 1) low-temperature formation in the
absence of hydrogen \citep{kroto:C60discovery}; 2) high-temperature ($T\ga 3000$\,K)
  formation in which case hydrogen may be present \citep{jag09}; 3) photo-chemical
processing of Hydrogenated Amorphous Carbon \citep[HAC,][]{sco97} ; and 4) destruction of PAHs \citep{Cami2011:IAU,ber12}. The plateaus are particularly interesting in this
  context since \citet{Stanghellini2007} attribute the 6--9\,$\mu$m and
10--13\,$\mu$m broad features in their sample of PNe to
unprocessed clusters of small carbonaceous grains and/or large PAH
clusters. \citet{Garcia2010:C60_PN_MW} on the other hand propose
that these are due to HACs, which they then suggest become
photo-chemically processed to produce the fullerenes.

\begin{figure*} 
\resizebox{\hsize}{!}{
\includegraphics[width=12cm]{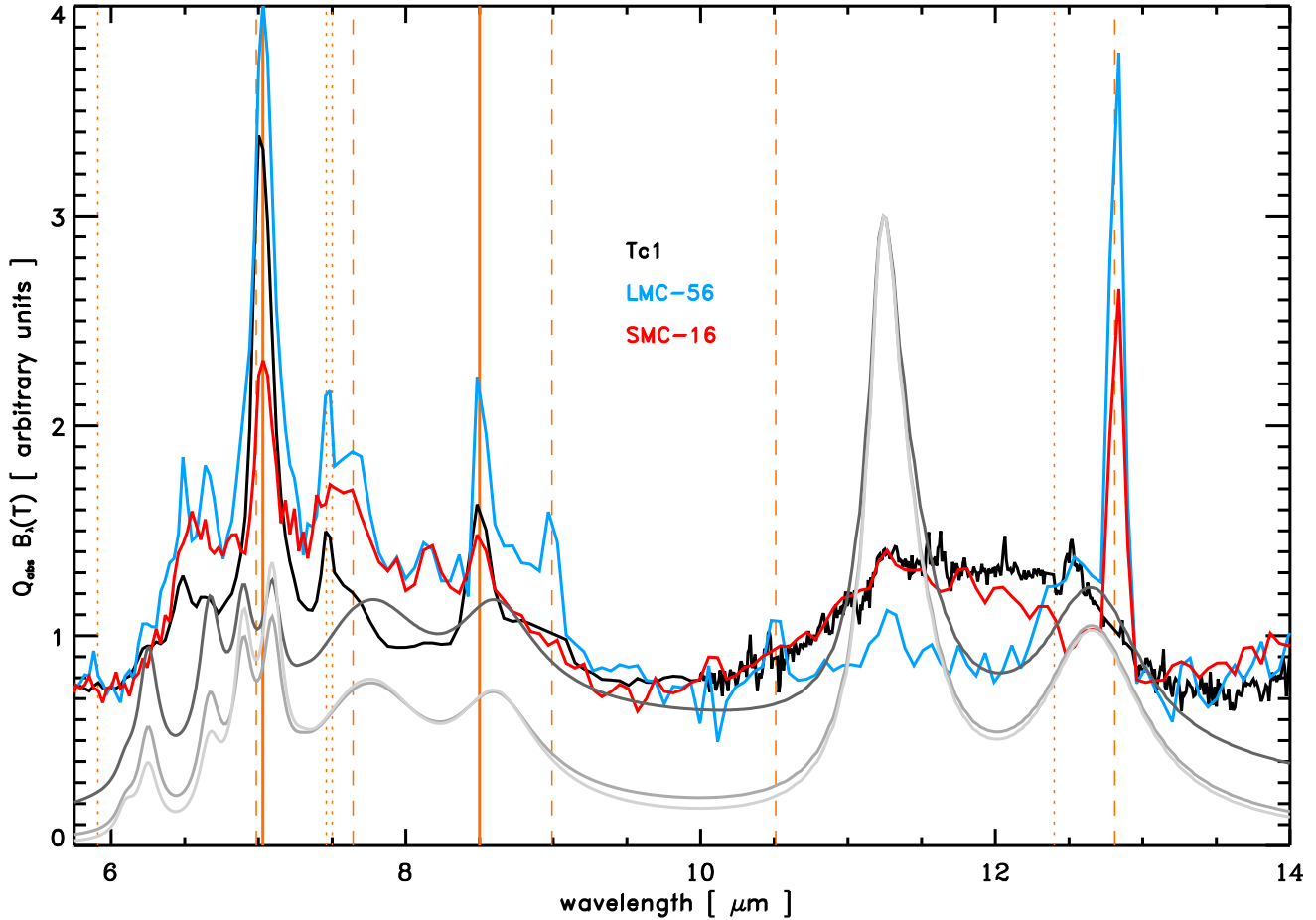}}
\caption{\label{Fig:specHAC} The continuum-subtracted spectra of the
  fullerene sources (Tc1, LMC\,56 and SMC\,16, black, blue and red
  lines, respectively compared to the absorption coefficient, $Q_{\rm abs}$
  multiplied by a blackbody, $B_\lambda(T)$, at $T =200$\,K, for
   3\,nm a-C:H particles with atom hydrogen fractions of  0.23 (dark-grey),
  0.29 (mid-grey) and 0.35 \citep[light-grey,][]{Jones:2012c}.  The vertical
  orange lines indicate the central positions of the fullerene bands
  (solid), gas phase ionic (dashed) and hydrogen lines (dotted).}
\end{figure*}

In Fig.~\ref{Fig:specHAC} we show the 6--14\,$\mu$m IR spectra of Tc1,
SMP~LMC~56 and SMP~SMC~16 compared to the IR spectra of
HAC/a-C:H particles with a radius of  3\,nm at a temperature of
  200\,K and with varying atom hydrogen fractions of 0.23, 0.29
and 0.35; this is equivalent to band gaps of 1.0, 1.25 and 1.5 eV,
respectively \citep{Jones:2012c}.  The spectra of these particles are
dominated by aliphatic and aromatic CC and CH bands \citep{kwo01,bus90}. As shown by \citet{Jones:2012b, Jones:2012c} these materials appear to also contain few olefinic
bands. The overlapping of the aliphatic, aromatic, and olefinic bands results in a plateau-like region
underlying the C$_{60}$ 7.0 and 8.5\,$\mu$m bands and also in
bands in the 11--13\,$\mu$m plateau  region\footnote{Longwards of 10$\mu$m the strength 
of the bands are not well characterized. The comparison of the observations and models in this region
 remains qualitative, particularly in terms of the band strengths.}. We note that the shape of
the continuum matches the observed 6 to 9\,$\mu$m plateau emission
under the fullerene bands and the atomic lines rather well in
these continuum-subtracted spectra. These spectral bands, underlying
the fullerene bands, appear to be rather invariant from source to
source, suggesting that the particles responsible for these
plateaus are therefore small enough to be stochastically
heated. However, there is a band at 6.5\,$\mu$m in the observed spectra
that we are unable to identify.

The spectra of Tc1, LMC56 and SMC16 appear to be consistent with
these H-rich carbonaceous materials (H atomic fraction 0.35 $-$
0.47).  Since exposure to UV radiation tends to reduce the
  hydrogen content of these HAC/a-C:H particles and aromatize the
  carbonaceous content, this indicates that the particles have not
been long-exposed to a strong radiation field. In the diffuse ISM,
where the dust is exposed to UV irradiation for millions of years, the
small HAC/a-C:H particles are expected to be H-poor, with an H atom
fraction of the order of only 0.05
\citep[e.g.,][]{Jones:2012a,Jones:2012b}.  Typical  PAHs (i.e. meaning strictly pure aromatics) can be ruled out based on the presence of aliphatic bands in the 6-9~$\mu$m spectral region and the shape of the 11-13~$\mu$m plateau.

One possible scenario that could therefore explain the particularity
of the fullerene sources is that small carbonaceous HAC/a-C:H
particles have recently emerged, either by being released from larger
(coagulated/accreted) particles or by the ablation of matter from a
denser medium, where the a-C:H materials were H-rich. Thus, the
carbonaceous dust in these sources would appear to be rather young, in
that it has not yet been exposed to stellar radiation for long enough
for it to be significantly dehydrogenated and aromatized. This
  could be understood in the context of optically thick dust not
  allowing the UV radiation to penetrate until very recently, on its way to becoming optically thin.
It is
perhaps these conditions that are conducive to fullerene molecule
formation by a top-down formation from much larger, 3D carbonaceous
particles \citep[e.g.,][]{Micelotta2012} rather
than via the PAH-warping mechanism proposed by \citet{ber12}.

As we saw in \S3 the central stars of the three PNe have similar (low) effective temperatures ($\sim$30\,000--45\,000~K). So far, fullerenes have been detected in low excitation environments (e.g. proto-PNe, reflection nebulae, YSOs). The link with low excitation environments has been noted by Evans et al 2012 in their study of R Coronae Borealis stars. 
As discussed in \citet{Micelotta2012}, fullerene formation is not favored in very low T$_{eff}$ sources because of the lack of sufficient UV photons to dehydrogenate and heat the dust. If it is too high however, fullerenes may not be formed because H-rich carbonaceous particles are destroyed before they can form fullerenes. Conversely, the fullerenes would be ionized in which case they have many more bands and their emission will be diluted making them much more difficult to detect. It is important to note that not all low excitation PNe show fullerene emission and so, other factors may play a role in the formation/destruction of fullerenes.

\section{Summary}

We have analyzed and compared the mid-IR spectrum of the three
fullerene-rich PNe Tc1, SMP~SMC~16, and SMP~LMC~56 to study the
  fullerene excitation conditions and formation mechanisms. The PNe
are low excitation and carbon-rich nebulae with low electron
temperature and thus are presumably young. While other factors
may be involved, these conditions seem to favor fullerene formation.

Spectroscopically, these unique PNe share many properties:
  they show the strongest and clearest circumstellar fullerene bands
(C$_{60}$) detected so far while showing little to no evidence for PAH
emission. In addition, they all have strong broad bands in the
6--9\,$\mu$m and 10--13\,$\mu$m range, a similar shape of the dust
continuum emission, and emission of the so called
30\,$\mu$m emission feature.  The observed fullerene band
strengths in the three sources is fairly similar (within
uncertainties) as well. However, the intensity of the radiation field
in these objects (as inferred by the fine-structure line ratio
[\ion{Ne}{3}]15.5\,$\mu$m/[\ion{Ne}{2}]12.8\,$\mu$m) varies by more than a
factor 10. Furthermore, the spatial profile of different dust
  components in Tc~1 indicates that the fullerene emission (C$_{60}$)
  peaks far away (6300-9700 AU) from the central source. All this is
  hard to reconcile with a thermal origin for the fullerene
  excitation, and thus favors fluorescence as the excitation
  mechanism.  Fluorescence of free, isolated
  C$_{60}$ molecules would be consistent with the observed band ratios
  if fluorescent emission from C$_{70}$
  contributes significantly to the 7.0\,$\mu$m band. Additional
  emitting materials could also be present and this needs to be further investigated.

It is likely that the observed broad bands at 6--9 and 10-13\,$\mu$m are related to the fullerene formation
process. We present model spectra of HAC particles
  of 3~nm and show that these can reproduce the 6-9\,$\mu$m band
with some degree of success, which may imply that fullerenes are
formed by photo-chemical processing of HAC.

The detection of C$_{60}$ and of other unidentified features in these
nebulae is very intriguing. Additional information and a larger
sample are required to discern the mechanisms that trigger fullerene
formation, and set their importance and role in the ISM, but it seems
that when the conditions are met, PNe can be important sources of
fullerenes.

\acknowledgments{J.B-S. wishes to acknowledge the support from a Marie Curie Intra-European Fellowship within the 7$^{th}$ European Community Framework Program under project number 272820.  JC, EP, EM acknowledge support from the  Natural Sciences and Engineering Research Council of Canada (NSERC)} 



\appendix

\section{A1. Flux Measurements}

We characterized the dust continuum by fitting a cubic spline
  through anchor points free of emission features along the spectrum
of each source (see Figure~1). We
have not attempted to define the continuum beyond $\sim$25\,$\mu$m
because this position marks approximately the onset of the 30\,$\mu$m
feature, which usually extends beyond 40\,$\mu$m. We
  quantified the amount of dust emission in each source by integrating
  the dust continua from 5--25\,$\mu$m (see
  Table~\ref{Table:Fluxes}). We then measured the fluxes in the
  fullerene bands and estimated the associated uncertainties in the
  following way.

In the high-resolution spectrum of Tc~1, the strong [\ion{S}{3}]
  line at 18.7\,$\mu$m is much narrower than the broad fullerene
  emission band, and it is thus straightforward to determine its
  contribution to the flux in the emission band. The fullerene flux
  can thus be measured by simply integrating over the entire band and
  subtracting this [\ion{S}{3}] contribution. The nominal uncertainty
  on the flux value obtained in this way is of the order of a
  percent. However, we feel that there is a larger source of
  uncertainty that originates from the positioning of the underlying
  continuum. To estimate this uncertainty, we fitted the observed band
  profile with three components: a broad Gaussian profile for the
  fullerene band\footnote{As was noted already in Paper I, the 18.9
   \,$\mu$m fullerene band is fairly symmetric, and can be well
    reproduced by a Gaussian profile. }, a narrow Gaussian profile for
  the [\ion{S}{3}] line, and a linear baseline that extends well
  beyond the band on both sides. In our best fit model, the uncertainty in the flux
  contained in this linear baseline is about 15\% of the flux
  contained in the broad Gaussian representing the fullerene band, and
  represents a much larger uncertainty on our flux measurements since
  it is not clear whether the flux in our baseline really belongs to
  the continuum, or to the fullerene band (which would imply the true
  fullerene band profile is not Gaussian). Our best estimate for the
  fullerene flux thus includes half of the flux in this baseline. We
  note that we also expect C$_{70}$ fullerenes to contribute to this
  band, and thus we subtracted the estimated C$_{70}$ contribution
  (see Paper I) to finally obtain the C$_{60}$ flux (see
  Table~\ref{Table:Fluxes}).

In the low-resolution observations of SMP~SMC~16 and SMP~LMC~56,
  the [\ion{S}{3}] line does contribute to the 18.9\,$\mu$m band, but
  it is much harder to separate from the fullerene band. Once again,
  we fitted the entire band using two Gaussian profiles and a linear
  baseline. This time though, we fixed the central wavelength of one
  component to the position of the [\ion{S}{3}] line and its width to
  the instrumental resolution. Again, we can reproduce the
  observations quite well, and we follow the same approach as for Tc~1
  to determine the fullerene flux and to estimate the uncertainty (see
  Table~\ref{Table:Fluxes}). \\

In the spectrum of Tc~1, the observed 17.4\,$\mu$m band is not as symmetric
  as the 18.9\,$\mu$m band, and shows some clear structure in its
  profile. Furthermore, there is weak emission in the red wing of the
  band amongst others due to a [\ion{P}{3}] line at 17.88\,$\mu$m,
  these however represent less than 10\% of the total flux in the
  band. We still used a Gaussian fit to the band to estimate the
  uncertainty introduced by mispossitioning the continuum; but since
  the band profile is asymmetric, we used the total integrated flux
  over the band (minus the contaminants and minus half the baseline
  flux) rather than the flux in the Gaussian fit as the best estimate
  for the fullerene flux. As before, we used the flux in the baseline
  as our uncertainty estimate. For SMP~SMC~16 and SMP~LMC~56, we could
  not estimate the contribution of the contaminants; therefore, the
  actual fullerene flux values are probably slightly lower than what
  we listed in Table~\ref{Table:Fluxes}; however, the contribution of
  these weak lines is probably smaller than the uncertainty on the
  flux value. 

The 8.5\,$\mu$m band was only observed at low-resolution in all
  three sources. There are no clear contaminants, and in all three
  sources, the band can be well reproduced by a Gaussian profile (with
  possibly a small red wing asymmetry in the spectra of SMP~SMC~16 and
  SMP~LMC~56). For Tc~1, even the error introduced by continuum
  effects is very small; for the other sources, a slight red wing
  asymmetry is the largest source of uncertainty (of the order of
  15\%). 

As was noted in Paper~I, the observed 7.0\,$\mu$m band in Tc~1 is
  a blend of C$_{60}$ emission and an unresolved [\ion{Ar}{2}] line
  that occurs at nearly the same wavelength. Thus, it was not clear
  what fraction of the emission band was due to C$_{60}$. Here, we
  tried the same approach as for the 18.9\,$\mu$m band in the
  low-resolution spectra of SMP SMC 16 and SMP LMC 56: we carried out
  a two-component Gauss fit, where the central wavelength of one
  component is fixed to the position of the [\ion{Ar}{2}] line and its
  width to the spectral resolution. In this way, we find that for
  Tc~1, most of the 7.0\,$\mu$m emission band is in fact due to
  fullerenes; only about 15\% of the flux is due to the [\ion{Ar}{2}]
  line. This agrees well with an estimate for the [\ion{Ar}{2}] flux
  obtained from the following abundance argument. \citet{Pottasch2011}
  compare Tc~1 to IC\,418, and find that these two PNe are very
  similar in terms of excitation properties and physical
  conditions. However, while the ionic abundance of Ar$^{++}$ is the
  same for both objects, that of Ar$^+$ is 5.8 times higher in Tc~1
  than in IC 418. Clearly, the discrepancy arises from the fact that
  the 7.0\,$\mu$m flux was interpreted as pure [\ion{Ar}{2}] emission
  and no contribution of C$_{60}$ was taken into account.  This then
implies that the flux measured in the 7.0\,$\mu$m band mostly
originates from the fullerene band, and only about 1/6 of the
emission is originating from the [\ion{Ar}{2}] line,
consistent with our measurements above. For SMP SMC 16, we similarly
find that the [\ion{Ar}{2}] flux is about 15\% of the total emission
in the band; for SMP LMC 56 however, the best fit solution has no
contribution of [\ion{Ar}{2}] (see Table~\ref{Table:Fluxes}). For
  this band, the uncertainty on the baseline is small, and the
  uncertainty on the flux values is dominated by the noise on the
  observations.


\end{document}